\pdfoutput=1
\documentclass[pre,twocolumn,aps,superscriptaddress,citeautoscript]{revtex4-2}

\usepackage[table]{xcolor}
\usepackage{booktabs}
\usepackage{color}
\usepackage{amsmath,amssymb}
\usepackage{graphicx}
\usepackage{footmisc}
\usepackage{bbold,soul}
\usepackage{lipsum}
\usepackage{enumitem}
\usepackage{hyperref}
\usepackage{physics}
\usepackage{gensymb}
\usepackage{watermark}

\usepackage{mc}

\begin{document}

\title{Accurate, uncertainty-aware classification of molecular chemical motifs from multi-modal X-ray absorption spectroscopy}

\author{Matthew R.~Carbone}
\email{mcarbone@bnl.gov}
\affiliation{Computational Science Initiative, Brookhaven National Laboratory, Upton, New York 11973, USA}

\author{Phillip M.~Maffettone}
\affiliation{National Synchroton Light Source II, Brookhaven National Laboratory, Upton, NY 11973, USA}

\author{Xiaohui Qu}
\affiliation{Center for Functional Nanomaterials, Brookhaven National Laboratory, Upton, New York 11973, USA}

\author{Shinjae Yoo}
\affiliation{Computational Science Initiative, Brookhaven National Laboratory, Upton, New York 11973, USA}

\author{Deyu Lu}
\email{dlu@bnl.gov}
\affiliation{Center for Functional Nanomaterials, Brookhaven National Laboratory, Upton, New York 11973, USA}

\date{\today}

\begin{abstract}
Accurate classification of molecular chemical motifs from experimental measurement is an important problem in molecular physics, chemistry and biology. In this work, we present neural network ensemble classifiers for predicting the presence (or lack thereof) of $41$ different chemical motifs on small molecules from simulated C, N and O K-edge X-ray absorption near-edge structure (XANES) spectra. Our classifiers not only reach a maximum average class-balanced accuracy of $0.99$ but also accurately quantify uncertainty. We also show that including multiple XANES modalities improves predictions notably on average, demonstrating a ``multi-modal advantage" over any single modality. In addition to structure refinement, our approach can be generalized for broad applications with molecular design pipelines.
\end{abstract}

\maketitle

\paragraph*{Introduction---}

Artificial intelligence and machine learning (AI/ML), and more broadly data-driven methods, have made a huge impact in our society over the last decade, with exciting applications in image processing, self-driving vehicles and natural language processing using generative AI. The adaptation of AI/ML methods in scientific research has quickly spread to a wide range of domains, from fundamental research in physics~\cite{carleo2017solving,nomura2017restricted,cai2018approximating,sturm2021predicting,miles2021machine,lee2023machine}, chemistry~\cite{Behler2007,gomez2018automatic,burger2020mobile}, biology~\cite{jumper2021highly} and materials science~\cite{carbone2019classification,torrisi2020random,rankine2022accurate,penfold2022deep} to applications in drug discovery pipelines~\cite{lavecchia2015machine,vamathevan2019applications} and autonomous experimentation/self-driving laboratories~\cite{epps2020artificial,stach2021autonomous,maffettone2021constrained,maffettone2021gaming,konstantinova2022machine,maffettone2023self}. 

In the field of atomic scale modeling, of particular interest are AI/ML approaches that allow for insights into the structure-property relationship, which is key to the understanding of the fundamental physics that ultimately leads to practical applications. These can take different forms, such as ML surrogates that predict molecular or materials properties from the atomic structure, thus bypassing expensive quantum mechanics simulations (i.e., the forward problem)~\cite{Behler2007,butler2018machine}, ML classifiers that extract physical descriptors from experimental measurements (i.e., the inverse problem)~\cite{Timoshenko2017,carbone2019classification}, or autoencoders that can interpret the physical meaning of latent space variables~\cite{routh2021latent,liang2023decoding}.

In this study, we will focus on a specific type of the inverse problem, which extracts local structural and chemical information from the near edge features of X-ray absorption spectra, known as X-ray absorption near edge structure (XANES). XANES is a widely used materials characterization technique that is element specific and sensitive to the local chemical environment of the absorbing site (e.g., charge state, coordination number and local symmetry). Recently, significant progress has been made in solving the inverse problem using ML techniques, including predicting the average coordination numbers of metal nanoparticles~\cite{Timoshenko2017,liu2021probing} and the local chemical environment of materials (e.g., transition metal cations~\cite{carbone2019classification,zheng2020random,torrisi2020random} and amorphous carbon~\cite{aarva2019understandingI,aarva2019understandingII}). In molecular systems, Carbone \emph{et~al.} performed principal component analysis (PCA) of simulated molecular N and O K-edge XANES and showed that the features in the reduced dimension has a strong correlation with the N and O-containing chemical motifs~\cite{carbone2020machine}. Tetef \emph{et al.} used various unsupervised methods (including PCA, variational autoencoders, t-distributed stochastic neighbour embedding, and the uniform manifold approximation and projection) on P and S K-edge XANES to quantify the spectral sensitivity on the local chemical environment of organophosphorus and organosulfur compounds~\cite{tetef2021unsupervised,tetef2022informed}. They also showed that valence-to-core X-ray emission spectra can provide similar chemical information in character and content~\cite{tetef2021unsupervised}. Despite this recent progress, an accurate and uncertainty-aware XANES-based ML classifier of a broad range of molecular chemical motifs has not yet been reported. 

In this letter, we show how uncertainty-aware neural network classifiers can be used to predict 41 different chemical motifs to very high accuracy, using multiple XANES signals as input. These ML classifiers can have important applications in environmental chemistry~\cite{myneni2002soft}, paleontology~\cite{boyce2002organic}, organic geochemistry~\cite{cody1996application}, and space science~\cite{uesugi2014sequential}. We also evaluate the effectiveness of the UQ algorithms, and demonstrate a ``multi-modal advantage": the accuracy of the classifiers is significantly higher when using multiple XANES signals from different absorbing elements when compared to using only single modalities independently.

\paragraph*{Methods---}

\begin{figure}[htbp!]
    \centering
    \includegraphics[width=1.0\columnwidth]{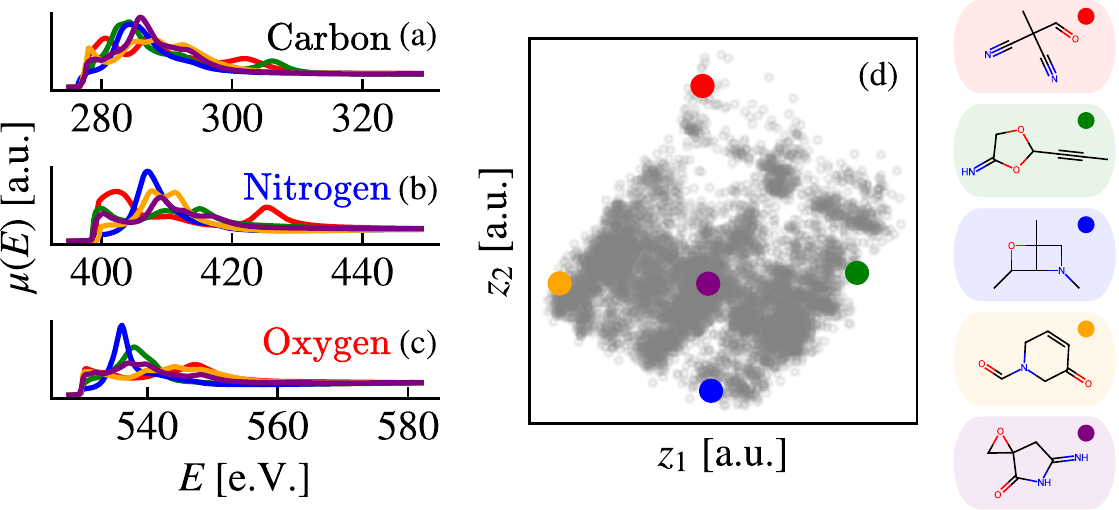}
    \caption{
    Five exemplary molecules from the $\mathcal{D}_\mathrm{CNO}$ database and their corresponding spectra. Carbon, nitrogen and oxygen K-edge XANES are shown (a-c). PCA is used to decompose the full spectral dataset (C, N and O K-edge XANES stitched together). The coefficients of the first two principal axes, $z_1$ and $z_2,$ are plotted in (d) as grey background. The location of these example spectra are shown as colored markers. The corresponding molecules are show to the right of (d).
    \label{fig:example-spectra}
    }
\end{figure}

\begin{figure}[htbp!]
    \centering
    \includegraphics[width=1.0\columnwidth]{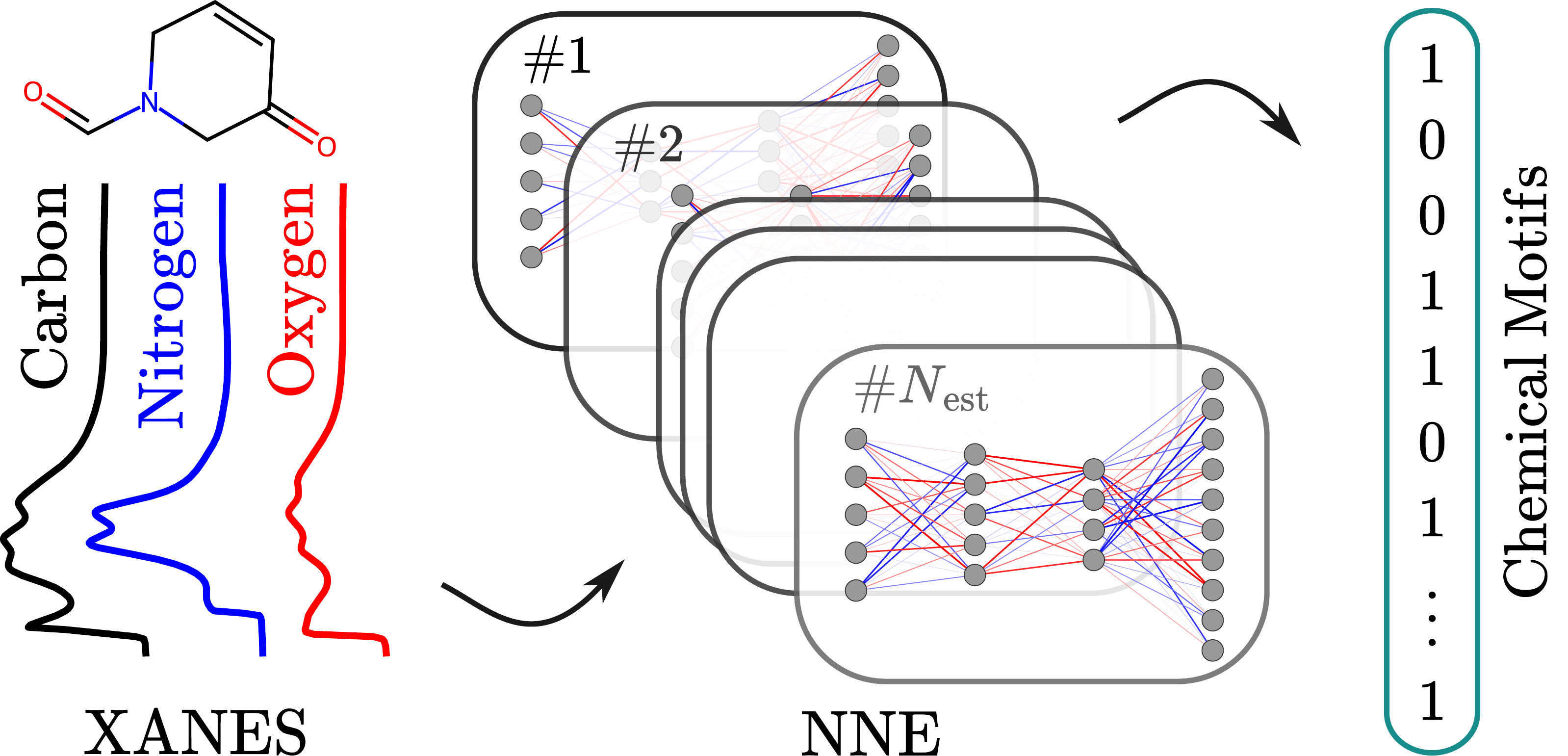}
    \caption{
    Multimodal XANES spectra (C, N and O signals) are fed into individual neural networks, which constitute a neural network ensemble (NNE) of $E$ estimators, to predict a binary vector, each entry of which corresponds to the presence (or lack thereof) of a specific chemical motif.
    \label{fig:workflow-cartoon}
    }
\end{figure}

To construct our dataset, molecular carbon, nitrogen and oxygen K-edge XANES were computed using the multiple scattering code FEFF9~\cite{rehr2010parameter}. Density Functional Theory-relaxed molecular structures are sourced from the QM9 database~\cite{ramakrishnan2014quantum}, a subset of the GDB-17 chemical universe~\cite{ruddigkeit2012enumeration}. Each molecule in the QM9 dataset ($\approx$134k small molecules) contains at most 9 heavy atoms (C, N, O and F). We partition the QM9 database into non-disjoint subsets based on which atoms are present (and thus which XANES signals exist). For example, $\mathcal{D}_\mathrm{CN}$ is the subset of QM9 in which every molecule contains \textit{at least} one C and one N atom, and thus at least a C and N XANES signal. We use the same database as in Ghose~\emph{et~al.}~\cite{ghose2023uncertainty}, which is available open access~\cite{carbone_matthew_r_2023_7554888}. As a preliminary analysis, we perform PCA on the $\mathcal{D}_\mathrm{CNO}$ dataset and show that spectra of exemplary molecules with different chemical motifs are well separated in the spectral latent space (see Fig.~\ref{fig:example-spectra}).

The trend from the PCA analysis agrees with the studies in the literature that established the spectral fingerprints (in C, O, and N K-edge XANES) of molecular chemical motifs~\cite{cody2008quantitative}. In C K-edge XANES, distinct lower energy features in alkenyl and aromatic species are assigned to 1s to $\pi^*$ transitions~\cite{cody1996application,urquhart2002trends} and relative intense near edge features in saturated carbon (e.g., methyl and methylene) are assigned to the transitions of 1s to the 3p/$\sigma^*$ hybrid state involving mixing of a 3p Rydberg state with a $\sigma^*$ C-H orbital~\cite{stohr2013nexafs}. Similarly, in N K-edge XANES, the 1s to $\pi^*$ transitions are characteristic of unsaturated N species (e.g., imine and nitrile) as compared to the 1s to the 3p/$\sigma^*$ excitations in saturated N chemical motifs (e.g., amine and pyrrole)~\cite{cody2008quantitative}. O K-edge XANES of small molecules can be used to extract ratio of single to double bonded oxygen~\cite{cody2008quantitative}.

As the goal of this work is inferring chemical information from XANES spectra, we choose 41 chemical motifs (see Fig.~\ref{fig:SI-chemical-motifs} and Table~\ref{tab:SI-chemical-motifs} in the Supplemental Material for their definitions), including many common functional groups, as the target quantity for machine learning. Using SMARTS pattern matching and Open Babel~\cite{o2011open}, each molecule is assigned a list of the chemical motifs (alcohols, ketones, aromatic character, alkynes, nitriles, etc.) that it contains. For each dataset, only chemical motifs that occur with frequency $\in [0.05, 0.95]$ in the dataset are used for training and evaluation.

The workflow of this work is shown in Fig.~\ref{fig:workflow-cartoon}. The QM9 XANES spectra, $\vecmu^{(i)},$ are used as inputs to fully connected deep neural networks (DNNs), which are tasked with predicting the presence (or lack thereof) of $41$ chemical motifs. Each modality of XANES spectra (C, N and O) is of length 200 on a pointwise grid. The target, $\vecy^{(i)},$ is a binary vector, where $y_j^{(i)} = 1$ if at least one instance of chemical motif $j$ is present on training example $i,$ and 0 otherwise (in principle, in experimental measurements the number of absorbing atoms can be associated with the edge jump~\cite{}, which would provide an additional constraint in e.g. the inverse problem). The last layer of each DNN is a sigmoid activation function, $\sigma(x) = 1/(1+e^{-x}) \in (0, 1),$ and a mean squared error loss is used to train the continuous neural network output, $\hat{\vecy}^{(i)},$ to best predict the binary targets. While this is formally a multi-task, binary \textit{classification} problem, models are trained using a regression approach and produce continuous outputs. Results are then rounded during inference.

For the main results of this work, each dataset is partitioned into a training (72.25\%), validation/hyper-parameter tuning (12.75\%) and testing (15\%) split. We perform hyper-parameter tuning using Optuna~\cite{akiba2019optuna}, in which the dropout rate, neural network architecture and initial learning rate are varied. All results in this manuscript are presented on the testing set using the best model as evaluated on the validation set.

To quantify the performance of our models, we utilize the class-balanced accuracy (CBA) scalar metric. The CBA is defined as the mean of the sensitivity and specificity,
$$\mathrm{CBA} = \frac{1}{2}\left[\frac{T_+}{T_+ + F_-} + \frac{T_-}{T_- + F_+} \right],$$
where $T_\pm$ ($F_\pm$) are the number of true (false) positives (negatives). In the case of a model taking advantage of a highly imbalanced data (one class dominates in propensity and the model predicts that class almost every time), the CBA will reduce to $0.5$ in the binary classification case (unlike the accuracy, which will be erroneously inflated). This is important for this problem, as some chemical motifs occur with frequency only $\approx 5\%.$

The CBA core obtained from the testing set can be treated as a proxy for how well the ML model will perform on unseen data, like in standard ML surrogate use cases. However, such models are not ``uncertainty aware", i.e., they do not produce a measure of confidence on any given training example. As ML methods are inherently interpolative procedures~\cite{mcPerspective}, knowing when data drifts out-of-sample is of key importance. Uncertainty quantification (UQ) is a set of statistical methods broadly classified as predictive methods for accurately estimating statistical uncertainty. UQ as applied to ML includes a measure of model confidence in addition to the model output. In this work, we apply neural network ensembles (NNEs, in similar fashion to previous research by the authors~\cite{ghose2023uncertainty}) to quantify the uncertainty in the model predictions. Twenty individual estimators are generated by a combination of training set down-sampling and architecture randomization. Starting from the scaffold of the best performing architecture from the previous hyper-parameter search, the number of neurons in each layer is randomly changed by adding or subtracting a small integer. In addition, a different random subset of the training set (5\%) is removed for each estimator in the ensemble. These two methods together create sufficient model diversity to leverage the ``wisdom of the crowd" principle~\cite{settles2009active}: the estimate of the aggregated opinion of a diverse set of estimators is more robust than any single estimator. Such NNEs are generally a more sensible option than any single estimator for production applications.

\paragraph*{Results/Discussion---}

\begin{figure*}[htbp!]
    \centering
    \includegraphics[width=2\columnwidth]{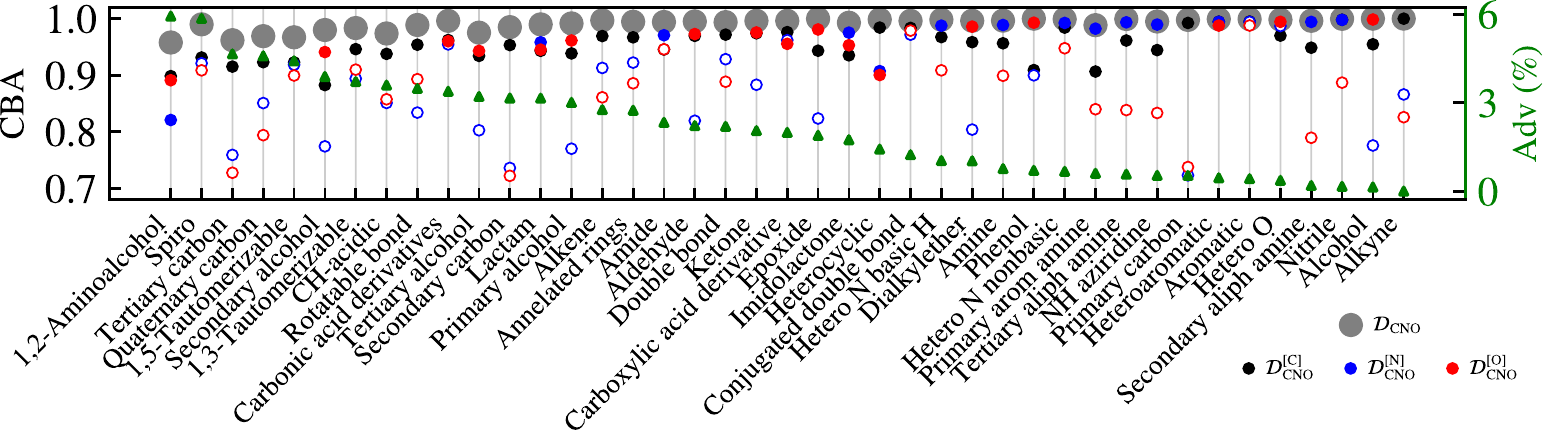}
    \caption{
    CBA scores for each model of the three single-modality models and the multi-modal model. Results are resolved by chemical motif and sorted (left to right) by decreasing multi-modal advantage (``Adv") as evaluated using the CBA metric. This advantage is defined by the difference of the $\mathcal{D}_\mathrm{CNO}$ result and the best performing single modality result. Open circles represent cases in which the XANES signal originates from at atom not contained in the corresponding chemical motif (e.g. an open blue circle for an alcohol).
    \label{fig:multi-modal-advantage}
    }
\end{figure*}

We first evaluate the CBA performance of single- and multi-modality models on the testing set. Using $\mathcal{D}_\mathrm{CNO},$ we execute the previously described pipeline (training, Optuna-powered hyper-parameter tuning, and evaluation on the testing set) on only certain subsets of the feature space. For example, we define $\mathcal{D}_\mathrm{CNO}^{[\mathrm{C}]}$ as the same dataset as $\mathcal{D}_\mathrm{CNO}$ (including the same train/validation/testing splits) but only using the C XANES as input. For each $\left\{\mathcal{D}_\mathrm{CNO}^{[\mathrm{C}]}, \mathcal{D}_\mathrm{CNO}^{[\mathrm{N}]}, \mathcal{D}_\mathrm{CNO}^{[\mathrm{O}]}, \mathcal{D}_\mathrm{CNO} \right\},$ an optimal model is found via hyper-parameter tuning, and the results of the best model evaluated on the testing set. These results are shown in Fig.~\ref{fig:multi-modal-advantage}. 

Including additional modalities usually improves model performance (as graded by CBA), and at worst simply does not improve over single modalities. For example, nitriles and amides have distinct fingerprints in both the C and N K-edge XANES, and similarly alcohols and carboxyl groups do in C and O K-edge XANES. We highlight that these results are obtained independently from pure data analysis approach without domain knowledge, and they reproduce the known correlation in the literature~\cite{cody2008quantitative}. These signals are all that are necessary to perform accurate classification and thus including extra modalities does not substantially improve CBA. On the other hand, 1,2-aminoalcohols are more complex motifs (containing multiple absorbing atom types, multiple types of chemical bonds, etc.) that are harder to resolve without inputs from all three modalities, and thus result in roughly a 5\% CBA gain over the highest-performing single modality.

The complementary information from the combination of multimodal XANES measurements can be understood using a tight-binding picture. For a qualitative discussion, we ignore many-body physics and map the atomic structure of a small molecule to a tight-binding Hamiltonian that contains on-site orbital energies and hopping integrals. To further simplify the problem, we take the projected desnity of states of the unoccupied bands under the quasi-particle picture as a proxy of the K-edge XANES spectra. The peak positions and their intensities of the spectrum correspond to eigenvalues and the weights of $p$ orbitals of the absorbing atoms. Solving the inverse problem with single modality requires one to deduce the matrix elements of the tight-binding Hamiltonian (e.g., on-site energies of atoms other than the absorber and hopping integrals). Multimodal XANES signals supply more complete information of the unoccupied molecular orbitals (i.e., with weights on multiple atom types) and reduce the dimension of the parameter space (i.e., by knowing on-site energies of multiple atom types). As a result, multimodal XANES analysis represents an advantage in solving the inverse problem.

Furthermore, we highlight that \textit{all} CNO models are incredibly accurate, with an average CBA score of $0.99 \pm 0.01$ across all examined chemical motifs, which is promising for future downstream applications such as targeted molecular refinement~\cite{gomez2018automatic}. We also find similar results using all combinations of $\mathcal{D}_{AB}$ for $A, B \in \{\mathrm{C}, \mathrm{N}, \mathrm{O}\}$ (which are distinct datasets from $\mathcal{D}_\mathrm{CNO}$), with the highest multi-modal advantage as large as 7\% in the case of $\mathcal{D}_\mathrm{NO}.$ Finally, there are certain cases where XANES signals originate from atoms not precisely contained in the given chemical motifs, such as a signal from the N K-edge trying to predict an alcohol. These cases are kept for completeness and shown as open circles in Fig.~\ref{fig:multi-modal-advantage}. Cases in which these signals contribute positively to the accuracy of the models are likely due to non-local correlations between different parts of the small molecules, which may worth further investigation. 

For future applications such as targeted molecular structure generation via structure search algorithms, it is important that the confidence of model prediction can be quantified. That way, an out-of-sample prediction can be confidently discarded, and confident predictions can be trusted. Towards this end, we perform UQ in similar fashion to Ref.~\onlinecite{ghose2023uncertainty}, using an ensemble of 20 models built off of the scaffold of the best performing model from the hyper-parameter tuning procedure. To evaluate the performance of the UQ procedure, we analyze the CBA as a function of the ``number of deviating estimators" (NDE): the number of estimators that deviate from the aggregate. Consider $N_\mathrm{est}$ total estimators, where $N_\mathrm{est} - n$ estimators predict $\hat{y}_j=0$ (or 1), and $n < N_\mathrm{est} - n$ estimators predict 1 (or 0). In this case, $n$ estimators have deviated from the majority opinion, and thus $n$ is the NDE. The NDE can also be regarded as an integral proxy for the spread of the estimator predictions. Regardless, as NDE increases, it is intuitive that CBA will decrease, since more disagreement would imply more uncertainty. Note that when $n = N_\mathrm{est}/2$, CBA will always be $1/2$ according to its definition (when all estimators disagree, the prediction will be $0.5,$ which is always rounded to 0. This means that only a single class is predicted, and in this case, CBA is equal to 0.5).

\begin{figure}[htbp!]
    \centering
    \includegraphics[width=\columnwidth]{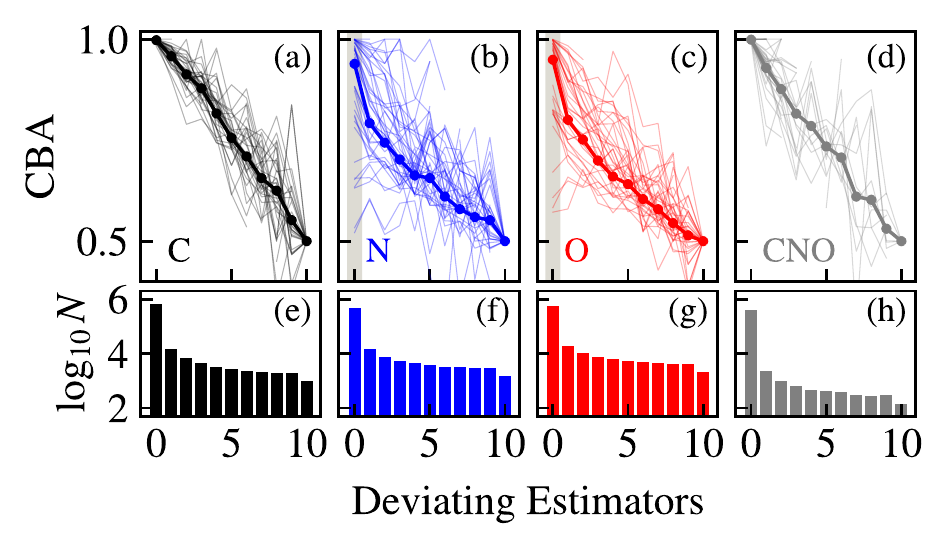}
    \caption{
    Uncertainty quantification using neural network ensembles, resolved by chemical motif. (a)-(d) Class-balanced accuracy as a function of the number of deviating estimators, for each chemical motif (thin lines) and averaged across the entire dataset. Data in which there are $<10$ examples are discarded from each bin due to lack of sufficient statistics. (e)-(h) The number of total training examples ($\times$ the number of chemical motifs) contained in the full dataset as a function of the number of deviating estimators.
    \label{fig:uq}
    }
\end{figure}

We plot the average testing set CBA results over the NNEs and chemical motifs in Fig.~\ref{fig:uq}(a)-(d) with the resolved individual chemical motifs shown as thin lines. For all three single modalities as well as the CNO multi-modal result, CBA decreases monotonically with NDE on average. For $\mathcal{D}_\mathrm{CNO}^{[\mathrm{C}]}$ and $\mathcal{D}_\mathrm{CNO},$ when all estimators are in agreement, the models are essentially perfectly accurate. These cases also constitute the vast majority of the data, as shown in Fig.~\ref{fig:uq}(e)-(h). As soon as a single estimator begins to deviate, the CBA score drops. The effect is even more significant in the $\mathcal{D}_\mathrm{CNO}^{[\mathrm{N}]}$ and $\mathcal{D}_\mathrm{CNO}^{[\mathrm{O}]}$ cases, where in contrast, the model CBA scores are much lower on average even with 0 deviating estimators. This is due to the chemical nature of chemical motifs in small molecules: almost all contain C as part of their backbone, making it logical that signals from N- and O-containing chemical motifs are embedded in C K-edge XANES. Conversely, there is no reason to expect that in general, N K-edge XANES contains a signature for O-containing chemical motifs (except due to dataset-dependent correlations, such as when N and O are often nearest-neighbors in a dataset). Indeed, almost all chemical motifs in Fig.~\ref{fig:uq}(b) and (c) with CBA less than $0.9$ at 0 NDE (corresponding to the indicated grey area) in the $\mathcal{D}_\mathrm{CNO}^{[\mathrm{N}]}$ and $\mathcal{D}_\mathrm{CNO}^{[\mathrm{O}]}$ results do not contain the absorbing element. While presented here for completeness, one would simply not use these models in real deployment scenarios in which high performance is demanded.

To further examine the advantage of utilizing information from multiple modalities in more realistic situations, we performed a secondary study in a subset of QM9 containing, where molecules with 8 or less heavy atoms were used for training/validation, and molecules with 9 heavy atoms were used for testing (henceforth referred to as the 8/9 split). This split tests two hypothesis: whether or not the multi-modal advantage is still present in a case with severely limited training data, and if these models can generalize to molecules slightly larger than those in the training set. We find that these models are also performant, with a CBA score of $0.96 \pm 0.03$ across all chemical motifs, and that the same UQ trends in Fig.~\ref{fig:uq} are recovered. These results are presented in detail in the Supplemental Material.


\paragraph*{Conclusions---}

We presented uncertainty-quantifying classifiers for predicting 41 different chemical motifs in small molecules from K-edge XANES spectra. Our models not only perform exceptionally well in the classification task, they are also able to accurately quantify uncertainty. Although the QM9 dataset represents a limited chemical space and simulated XANES can have discrepancies with experiment~\cite{carbone2019classification}, our workflow is entirely general and can in principle be applied to different levels of theory, more complex datasets, and experiment, if a large number of labeled measurements are available. In summary, we demonstrate the usefulness of uncertainty-quantifying models and multi-modal inputs and believe that both of them key to future progress in the field.

\paragraph*{Acknowledgements---} The authors would like to acknowledge helpful discussions with Mark S. Hybertsen. This material is based upon work supported by the U.S. Department of Energy, Office of Science, Office of Basic Energy Sciences, under Award Numbers FWP PS-030 and DE-SC-0012704. This research used the Theory and Computation resources of the Center for Functional Nanomaterials, which is a U.S. DOE Office of Science Facility, and the Scientific Data and Computing Center at Brookhaven National Laboratory under Contract No. DE-SC0012704.

\paragraph*{Software and data availability---} All code/data used in this manuscript can be found open source/access at 
\begin{itemize}
    \item \href{https://github.com/AI-multimodal/multimodal-molecules}{github.com/AI-multimodal/multimodal-molecules},
    \item \href{https://github.com/matthewcarbone/Crescendo}{github.com/matthewcarbone/Crescendo},
    \item \href{https://doi.org/10.5281/zenodo.8087823}{doi.org/10.5281/zenodo.8087823}.
\end{itemize}





\clearpage
\pagebreak
\widetext
\begin{center}
\textbf{\large Supplemental Material: \\ Accurate, uncertainty-aware classification of molecular chemical motifs from multi-modal X-ray absorption spectroscopy}
\end{center}

\setcounter{equation}{0}
\setcounter{section}{0}
\setcounter{figure}{0}
\setcounter{table}{0}
\setcounter{page}{1}
\makeatletter
\renewcommand{\thesection}{S\arabic{section}}
\renewcommand{\thesubsection}{S\arabic{section}.\arabic{subsection}}
\renewcommand{\theequation}{S\arabic{equation}}
\renewcommand{\thefigure}{S\arabic{figure}}
\renewcommand{\thetable}{S\arabic{table}}

\section{Methodological details}

We henceforth outline the details of our methodology, starting from the raw data and ending with the training and evaluation procedures.

We begin with the same dataset from Ref.~\onlinecite{ghose2023uncertainty}, which can be found at \href{https://doi.org/10.5281/zenodo.7554888}{doi.org/10.5281/zenodo.7554888}. This dataset contains the site-wise XANES spectra for all C, N and O sites in the QM9 database. From there, it is processed into molecular XANES spectra. Any molecule in which at least one XANES spectrum was missing/failed to converge was discarded. This new dataset can be found here: \href{https://doi.org/10.5281/zenodo.8087823}{doi.org/10.5281/zenodo.8087823}. The following steps, which are also contained in the \code{scripts} directory in our \code{multimodal-molecules} GitHub repository (all paths are relative to the root directory of this repository), are as follows.

\subsection{Identification of chemical motifs from SMILES representation}

From the file \code{xanes.pkl}, all SMILES strings are extracted and saved as a text file. Open Babel is then used in conjunction with the FP4 fingerprinting method, corresponding to SMARTS queries stored in the Open Babel GitHub repository \href{https://github.com/openbabel/openbabel/blob/master/data/SMARTS_InteLigand.txt}{github.com/openbabel/openbabel/blob/master/data/SMARTS\_InteLigand.txt}. Ultimately, this process results in a source of truth file \code{index.csv}, which contains a binary indicator of which functional groups are present on which molecule. Similarly, \code{xanes.pkl} is the mapping between SMILES and XANES spectra. These two files are used in constructing the training/validation/testing datasets in the next steps. These steps can be reproduced by running \code{scripts/00\_get\_functional\_groups.py} and \code{scripts/01\_create\_index.py} in succession.

\subsection{Dataset construction}

For every possible modality combination (and for the 8/9 dataset split), we create training/validation/testing splits. The data are saved in a format readable by our ``easy execution" machine learning code, Crescendo, which leverages Lightning and Hydra for easily trainable models and neural network ensembles. This format essentially requires a directory with NumPy binary files like \code{X\_train.npy}, \code{Y\_val.npy}, etc. A file \code{splits.json}, which indicates which training indexes to downsample during ensemble training, is also saved. These steps can be reproduced by running \code{scripts/02\_construct\_datasets.py}.

\subsection{Hyperparameter tuning setup}

Hyperparameter tuning was performed for all possible modality combinations using Hydra's Optuna plugin. Wrapper code in Crescendo was used to run these scripts. The following hyper-parameters were tuned in all cases:
\begin{itemize}
    \item Neural network architecture: starting from the input size of 200, 400 or 600 input neurons (the discretized, possibly multi-modal XANES spectrum), and ending with the output size corresponding to the number of chemical motifs in that dataset (between 36 and 42), the number of hidden layers is varied. The number of neurons per layer is determiend by a simple linear ramp from the input size to the output size. For example, a linear ramp of 2 hidden neurons between input size of 10 and output size of 4 would simply result in an architecture of $[10, 8, 6, 4].$ In these experiments, the number of hidden layers was chosen from the set $\{6, 9, 12, 15, 18, 21\}.$
    \item The training mini-batch size was chosen from the set $\{128, 256, 512, 1024\}.$
    \item The initial Adam optimizer learning rate was chosen from a continuous range $(0.001, 0.01).$
    \item The dropout rate was chosen from the continuous range $(0, 0.1).$
\end{itemize}

Optuna executed 100 total trials in order to determine the optimal hyper-parameter set. Hyper-parameter tuning was performed for 500 epochs, using an early stopping criterion and learning-rate-decrease-on-validation-loss-plateau scheduler. All models use batch normalization, except the last layer, and a Sigmoid output layer. The best set of hyper-parameters in each case was used as the starting point for the training of the neural network ensembles.

\subsection{Neural network ensemble details}

Starting from the hyper-parameters found in the previous section, a set of 20 estimators is each trained for at most 1000 epochs using the same criteria as in the hyper-parameter tuning. Each estimator has a randomized architecture, which is constructed by starting from the scaffold found in the hyper-paramter search, and randomly distorting each layer by rounded Gaussian noise with standard deviation 10. Further model diversity was created by a random down-sampling of the training sets, such that each estimator is trained on a different 95\% sampling of the full training set.

\subsection{Evaluation details}

Each individual model, whether part of an ensemble or not, is treated as its own \textit{classifier}. Thus, for a single estimator, its prediction is the binary vector corresponding to the rounded neural network output. In an ensemble, the prediction is the rounded mean of these binary vectors. The number of deviating estimators is of course the number of estimators that disagree with the majority opinion, as outlined in the main text.


\section{Supplemental Figures and Tables}

In this section, we present supplemental figures and tables to add additional context to the content presented in the main text.

\begin{figure}[htbp!]
    \centering
    \includegraphics[width=\columnwidth]{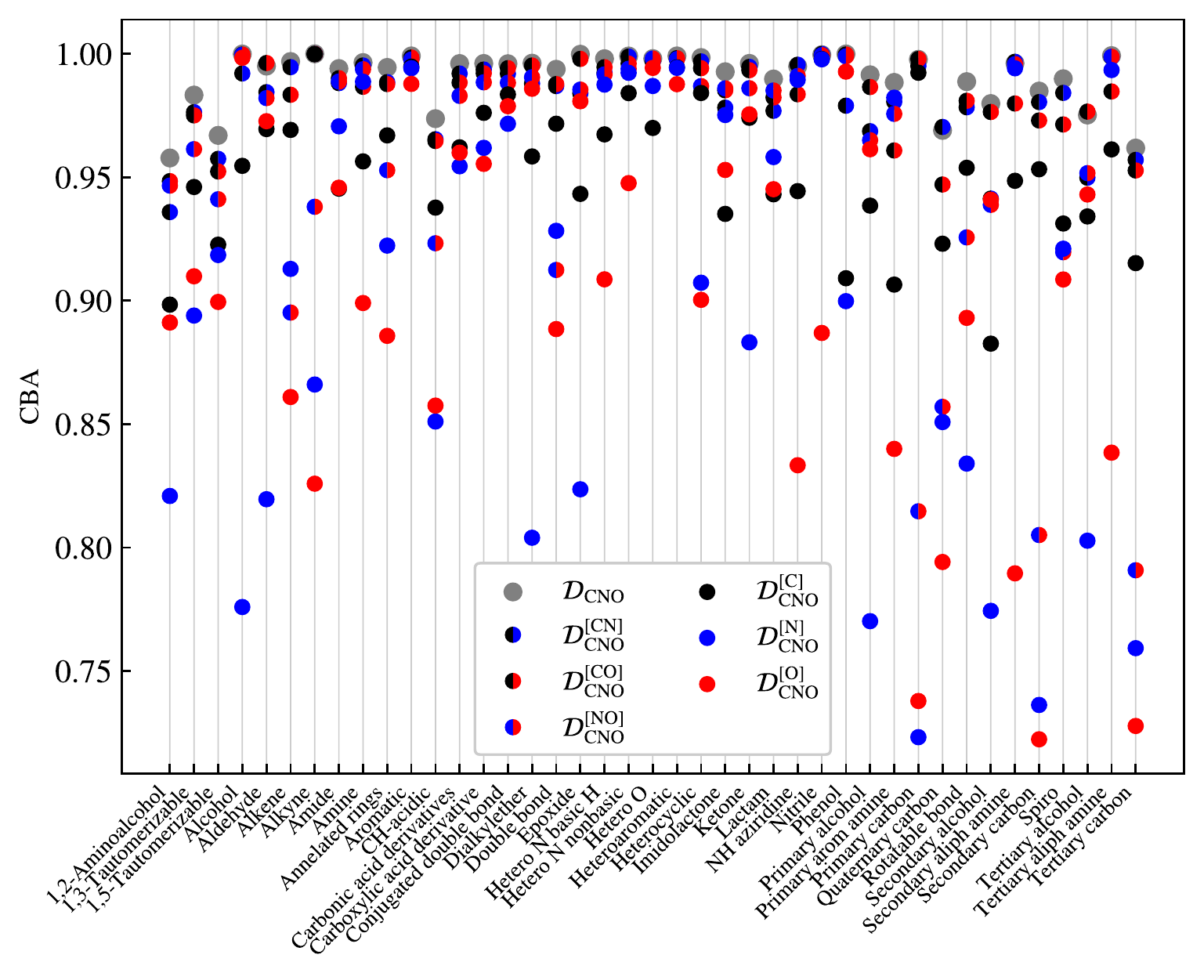}
    \caption{
        Demonstration of the multimodal advantage using pairs of modalities in addition to those presented in Fig.~\ref{fig:multi-modal-advantage}.
    \label{fig:SI-multi-modal-advantage}
    }
\end{figure}

First, we showcase a demonstration of the multimodal advantage when using \textit{pairs} of modalities (as opposed to all of the modalities as in the main text). The objective of this analysis is to confirm that pairs of modalities also provide an advantage over single modalities in most cases. These results are presented in Fig.~\ref{fig:SI-multi-modal-advantage}. Indeed, it is the case that the CBA score on average for double-modality models is higher than the corresponding single modalities. There are cases in which this is not the case (e.g. alkenes and the N- and O-containing spectra), but these correspond to situations where the XANES signal should not be assumed to contain the relevant information required to make the predictions.

\begin{figure}[htbp!]
    \centering
    \includegraphics[width=0.5\columnwidth]{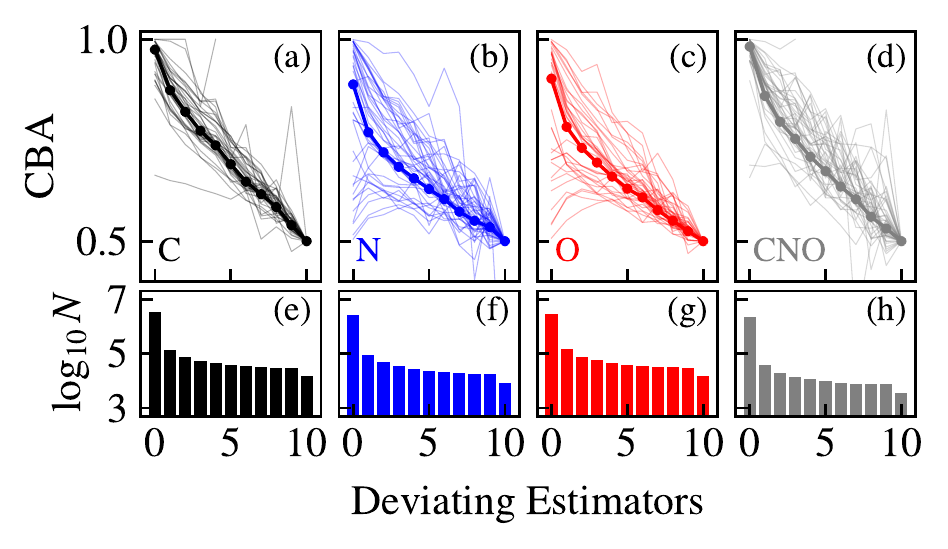}
    \caption{
    Uncertainty quantification using neural network ensembles for the 8/9 split datasets. Notation is identical to Fig.~\ref{fig:uq}.
    \label{fig:SI-uq-cutoff8}
    }
\end{figure}

Next, in Fig.~\ref{fig:SI-uq-cutoff8}, we show the UQ results of the 8/9 split dataset. This dataset used only molecules with up to and including 8 heavy atoms/molecule for training and validation, but tested on the (far larger) split of molecules containing 9 heavy atoms/molecule. While the models perform slightly worse overall, they are still quite robust and provide accurate UQ in almost all cases, mirroring the results in the main text using the random data splits.

\begin{figure}[htbp!]
    \centering
    \includegraphics[width=\columnwidth]{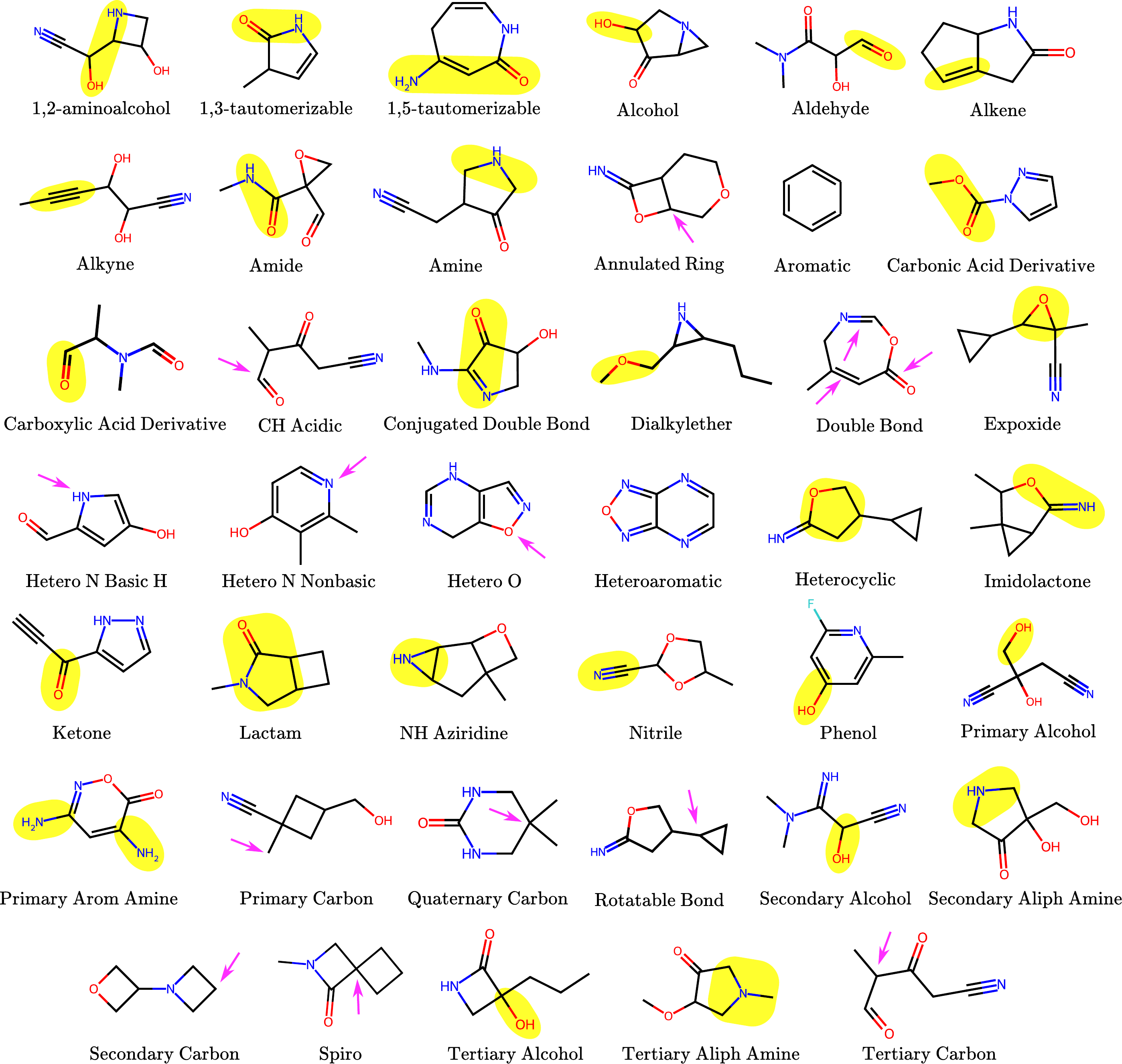}
    \caption{
        Enumerated chemical motifs used in Fig.~\ref{fig:multi-modal-advantage} of the main text. Details of each chemical motif are presented in Tab.~\ref{tab:SI-chemical-motifs}.
    \label{fig:SI-chemical-motifs}
    }
\end{figure}

\begin{table*}[!htb]
\caption{\label{tab:SI-chemical-motifs}%
Further details on the chemical motifs. Note that some of these classifications are slight misnomers. For example, a phenol group is formally of the chemical formula C$_6$H$_5$OH, but the FP4 fingerprinting method includes alcohols bonded to \emph{any} aromatic group. A ``+" symbol is used in place of a triple bond.
}
\begin{ruledtabular}
\begin{tabular}{lll}
\textrm{Chemical Motif} & \textrm{Representation} & \textrm{Description} \\
\colrule
\text{1,2-aminoalcohol} & - & An amine and alcohol separated by two carbon atoms \\
\text{1,3-tautomerizable} & - & Structural isomers that interconvert at the 1 and 3 positions \\
\text{1,5-tautomerizable} & - & Structural isomers that interconvert at the 1 and 5 positions \\
\text{Alcohol} & COH & A carbon single-bonded to an oxygen atom \\
\text{Aldehyde} & RCH=O & A C-O double bond where C is attached to at least one H \\
\text{Alkene} & $\mathrm{R}$$\mathrm{R}'$C=C$\mathrm{R}''$$\mathrm{R}'''$ & A C-C double bond \\
\text{Alkyne} & RC+C$\mathrm{R}'$ & A C-C triple bond \\
\text{Amide} & RC(=O)N$\mathrm{R}'$$\mathrm{R}''$ & A carbon double-bonded to an oxygen and single bonded to a nitrogen \\
\text{Amine} & R$\mathrm{R}'$$\mathrm{R}''$N & A nitrogen bonded to at least one carbon chain \\
\text{Annulated Ring} & - & Fused rings such that more than one rings shares at least two atoms \\
\text{Aromatic} & - & A conjugated system containing $4n+2$ $\pi$-electrons \\
\text{Carbonic Acid Derivative} & - & A molecule derived from a carbonic acid \\
\text{Carboxylic Acid Derivative} & - & A molecule derived from a carboxylic acid \\
\text{CH Acidic} & - & C-H bonded to a carbon on a carbonyl, nitro, or similar \\
\text{Conjugated Double Bond} & - & A double bond that is part of an extended $\pi$ orbital system \\
\text{Dialkylether} & RO$\mathrm{R}'$ & An oxygen bonded to two carbon chains \\
\text{Double Bond} & - & A molecule containing a double bond \\
\text{Epoxide} & - & An ether contained in a 3-membered ring of carbon atoms \\
\text{Hetero N Basic H} & - & A nitrogen contained in a ring with a basic H \\
\text{Hetero N Nonbasic} & - & A nitrogen contained in a ring which does not have basic character \\
\text{Hetero O} & - & Oxygen atom contained in a ring \\
\text{Heteroaromatic} & - & A non-carbon atom contained in an aromatic system \\
\text{Heterocyclic} & - & A non-carbon atom contained in a ring \\
\text{Imidolactone} & - & A carbon atom double-bonded to a nitrogen and single-bonded to an oxygen \\
\text{Ketone} & R(C=O)$\mathrm{R}'$ & A carbon atom double bonded to an oxygen and two carbon chains \\
\text{Lactam} & - & A cyclic amide \\
\text{NH Aziridine} & - & An amide contained in a 3-membered ring of carbon atoms \\
\text{Nitrile} & RC+N & A C-N triple bond \\
\text{Phenol} & - & An alcohol connected to an aromatic ring \\
\text{Primary Alcohol} & RCH$_2$OH & An alcohol in which the base carbon is bonded to two hydrogen atoms \\
\text{Primary Arom Amine} & - & An amine group connected to an aromatic ring \\
\text{Primary Carbon} & RCH$_3$ & A carbon atom bonded to three hydrogen atoms \\
\text{Quaternary Carbon} & R$\mathrm{R}'$$\mathrm{R}''$$\mathrm{R}'''$C & A carbon atom bonded to four carbon atoms \\
\text{Rotatable Bond} & - & A rotatable bond of $\sigma$ character \\
\text{Secondary Alcohol} & R$\mathrm{R}'$CHOH & An alcohol in which the base carbon is bonded to one hydrogen atom\\
\text{Secondary Aliph Amine} & R$\mathrm{R}'$NH & An amine bonded to two carbon groups\\
\text{Secondary Carbon} & R$\mathrm{R}'$CH & A carbon atom bonded to two carbon atoms\\
\text{Spiro} & - & Two rings that share a single atom\\
\text{Tertiary Alcohol} & R$\mathrm{R}'$$\mathrm{R}''$COH & An alcohol in which the base carbon is bonded to three carbon atoms\\
\text{Tertiary Aliph Amine} & R$\mathrm{R}'$$\mathrm{R}''$N & An amine bonded to three carbon groups\\
\text{Tertiary Carbon} & R$\mathrm{R}'$$\mathrm{R}''$C & A carbon atom bonded to three carbon atoms
\end{tabular}
\end{ruledtabular}
\end{table*}

Finally, we present some additional information on some of the commonly occurring chemical motifs in the $\mathcal{D}_\mathrm{CNO}$ dataset. Fig.~\ref{fig:SI-chemical-motifs} showcases examples of these motifs, and Tab.~\ref{tab:SI-chemical-motifs} contains additional details on each chemical motif. For further details, we refer the reader to the soure of truth for the SMARTS patterns which can be found in the previously mentioned GitHub repository, \href{https://github.com/openbabel/openbabel/blob/master/data/SMARTS_InteLigand.txt}{github.com/openbabel/openbabel/blob/master/data/SMARTS\_InteLigand.txt}.

\clearpage
\pagebreak

%

\end{document}